\documentclass{article}\def\address#1{}\def\abstracts#1{\small\noindent#1}

\usepackage{amsmath}
\usepackage{amsfonts}
\usepackage{amssymb}
\usepackage{epsfig}

\newtheorem{lemma}{Lemma}
\newtheorem{defi}{Definition}

\def\qed{\leavevmode\unskip\penalty9999 \hbox{}\nobreak\hfill
     \quad\hbox{\leavevmode  \hbox to.77778em{%
               \hfil\vrule   \vbox to.675em%
               {\hrule width.6em\vfil\hrule}\vrule\hfil}}}

\def\HH{{\cal H}}\def\MM{{\cal M}}
\def\BB{{\cal B}}
\def\LL#1{{\cal L}^{#1}}
\def\idty{{\leavevmode{\rm 1\ifmmode\mkern -5.4mu\else\kern -.3em\fi I}}}
\def\tr{{\rm tr}}
\def\norm#1{\left\Vert{#1}\right\Vert}
\def\abs#1{\left\vert{#1}\right\vert}

\def\Rl{{\mathbb{R}}}

\def\Qobs{{\bf Q}}
\def\Pobs{{\bf P}}
\def\Mobs{{\bf M}}\def\Mdens{{\bf m}}\def\Mobsc{\Mobs^{\rm av}}
\def\Fobs{{\bf F}}
\def\degf{n} % #degrees of freedom
\def\Ccs{{\cal C}_{00}} % fcs of compact support
\def\Cuc{{\cal C}_{\rm uc}}
\def\parity{\lower.8pt\hbox{$\Pi$}}

\begin{document}
%* Top Matter
\title{The Uncertainty Relation for Joint Measurement of Position and Momentum}
\author{R.~F. Werner}
\address{Institute for Mathematical Physics, TU Braunschweig,\\
Mendelssohnstra{\ss}e 3,
  D-38106 Braunschweig, Germany \\Email: {r.werner@tu-bs.de}}
 \maketitle
 \abstracts{We prove an uncertainty relation, which imposes a bound
 on any joint measurement of position and momentum. It is of the
 form $(\Delta P)(\Delta Q)\geq C\hbar$, where the `uncertainties'
 quantify the difference between the marginals of the joint
 measurement and the corresponding ideal observable. Applied to an
 approximate position measurement followed by a momentum
 measurement, the uncertainties become the precision $\Delta Q$ of
 the position measurement, and the perturbation $\Delta P$ of the
 conjugate variable introduced by such a measurement.
 We also determine the best constant
 $C$, which is attained for a unique phase space covariant measurement.}

 \vskip20pt{\sc Dedicated to Alexander S. Holevo on the occasion of his 60$^{\rm
 th}$ birthday}

%%%%%%%%%%%%%%%%%%%%%%%%%%%%%%%%%%%%%%%%%%%%%%%%%%%%%%%%%%%%%%%%%%%%%%
\section{Introduction}

Heisenberg's Uncertainty Relation
 $(\Delta Q)(\Delta P)\geq\hbar/2$ is
one of the most fundamental features of quantum theory, and is
taught in even the most basic course on the subject. All too
often, however, teachers succumb to the persistent bad habit of
proving the relations as an inequality on variances for arbitrary
state preparations, but then to go on to explain their `physical
meaning' in terms of a perturbation of the momentum of a particle
caused by an approximate position measurement. Since the usual
proof contains nothing of that sort, attentive students quickly
get the impression that quantum uncertainty rubs off on their
teachers as some kind of conceptual fuzziness. Our aim in this
paper is to state the measurement aspect of uncertainty as
rigorously as has become standard for the preparation aspect and,
of course, to prove the corresponding inequality.

Both aspects of uncertainty go back all the way to Heisenberg's
paper\cite{Hei} in which the relations were first introduced, and
it is perhaps instructive to disentangle the richness of
Heisenberg's paper a little bit. He begins his discussion with the
famous example of a position measurement on an electron by
observation under a $\gamma$-ray microscope: the resolution
$\Delta Q$ of such a device is of the order of the wavelength
$\lambda$ of the photons. However, the interaction gives the
electron a Compton kick, transferring an uncontrolled momentum of
the order of the momentum of the photon, i.e., $\Delta P\sim
2\pi\hbar/(\Delta Q)$. Heisenberg paraphrases this by saying that
precisely at the moment of interaction, i.e., at the moment the
electron's position ``becomes known'', the momentum ``becomes
unknown'' in accordance with the Uncertainty Relation. He observes
that this is related to the commutation relations, and announces
that this ``direct mathematical connection'' will be demonstrated
later in the paper. Disappointingly, this demonstration (on p.
180) turns out to be an order-of-magnitude discussion of the
spread of Gaussian wave packets.

\begin{figure}[ht]\label{boxes1}\centering
 \epsfxsize=4cm \epsffile{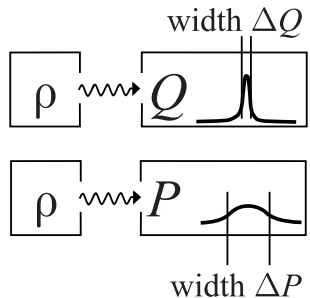}
 \caption{{\it The Preparation Uncertainty Relation refers to
    the variances in two separate ideal measurements on the same state.}}
 \end{figure}

After Heisenberg the stringency of this demonstration was improved
considerably, beginning with Kennard\cite{Ken}, and a version for general
non-commuting quantities by Robertson\cite{Rob}. These mathematical
formulations fix the meaning of $\Delta P$ and $\Delta Q$ as the square
root of variances, and replace Heisenberg's own notation ``$\sim$'' for
``of the order of magnitude of \ldots'' by a rigorous inequality, in
which even the constant $\hbar/2$ is precisely optimal. Since the proof
involves just an elementary application of the Schwarz inequality, it has
become standard textbook material, and Heisenberg himself seems to have
adopted it as the principal formulation of uncertainty in his later
writings. The meaning of the Uncertainty Relations in this formulation is
again summarized in Fig.~\ref{boxes1}. Obviously, they refer to two
separate experiments, in the sense that to each single quantum particle
either a position or a momentum measurement is applied. The preparation
is the same in both cases, so the relations are best seen as a constraint
on the possibility of {\it preparing} states with low variances.

But what became of the microscope? Clearly, Heisenberg discusses a
simple {\it measurement} process, in which the initial preparation
of the electrons plays no important role. Position and momentum
are both measured for the same particle (even if imperfectly). The
key observation is that the measurement of position necessarily
disturbs the particle, so that the momentum is changed by the
measurement. Indeed, it is a fundamental theorem of quantum theory
that {\it there is no measurement without perturbation}. More
precisely, if the output quantum states of a measuring device
coincide with the input states for all inputs, then the measured
values are statistically independent of the input, i.e., no
information is gained from the `measurement'. But this statement
captures none of the quantitative content of Heisenberg's
discussion.
 \begin{figure}[h]\label{boxes2}\centering
 \epsfxsize=7cm \epsffile{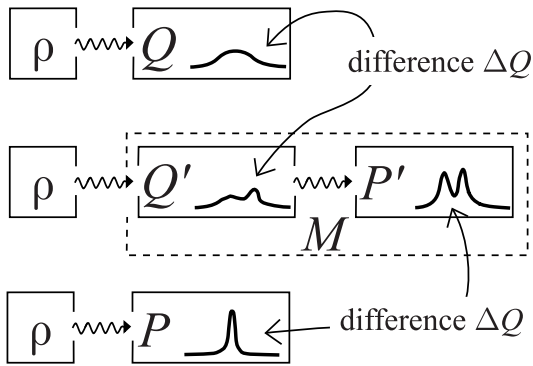}
 \caption{{\it The Measurement Uncertainty Relations studied in this paper
 refers to the deviations of the marginals of a joint measurement $\Mobs$ from
 the ideal position and momentum observables. The joint measurement can
 be realized by first a position measurement $\Qobs'$ and then a momentum
 measurement $\Pobs'$.
 }}
 \end{figure}
Figure~2 shows how we might understand the uncertainties for the
microscope: The quantum system is first subject to an approximate
position measurement $Q'$. This is not an ideal measurement $Q$, since
the $\gamma$-rays have non-zero wavelength. So $\Delta Q$ is some measure
of the difference between $Q$ and $Q'$. The next step is a measurement of
momentum. Due to the previous perturbation of the system, we cannot hope
to recover precisely the momentum of the initial particle. So if $P'$ is
the momentum measurement (including the prior perturbation) we will see a
difference $\Delta P$ to an ideal momentum measurement $P$. The claim of
the {\it Measurement Uncertainty Relations} is that $\Delta Q\Delta P\geq
C\hbar$ for some constant $C$. The aim of this paper is to do for this
relation what Kennard did for the Preparation Uncertainty Relation: to
give a rigorous definition of the quantities involved and to prove the
inequality as a Theorem.

The reason why this was not done 70 years ago might be that
$\Delta Q$ is a difference between observables like $Q$ and $Q'$,
which are never measured in the same experiment. Therefore a
quantity like the expectation of $(q-q')^2$ makes no sense at all.
So we have to define $\Delta Q$ as a distance between the
probability distributions of $q$ and $q'$, which requires some
conceptual work (see Section~\ref{s:probdiff}).

It is clear that we can make devices $M$, which for a particular
input state $\rho$ produce outputs with precisely the same
distributions as the ideal measurements. Indeed, we can simply
make $M$ a random generator for an arbitrary pair of
distributions. Such a device $M$ would utterly fail to reproduce
the distributions for other input states, of course. Therefore we
will define $\Delta Q$ (and similarly $\Delta P$) as the {\it
worst case} distances between the probability distributions of $Q$
and $Q'$.

In previous work on measurement uncertainty only covariant joint
measurements were considered, i.e., measurements with the expected
transformation behavior with respect to phase space translations.
In this case, which will also play a crucial role in the present
paper, the conceptual problem of interpreting the $\Delta$s is
much easier: in that case the marginals of a joint measurement can
be simulated by adding to the results of an ideal measurement some
noise, which is independent of the input. Hence any parameter
quantifying the size of the noise (e.g., the variance) will do.
Discussions of uncertainty in this setting can be found in many
places, not least in the work of Holevo\cite{HolevoUncert}. The
new contribution in this article is a definition of the $\Delta$s,
which makes sense without covariance, and the correspondingly
extended inequality.

This paper is organized as follows. We will describe the precise
definitions of $\Delta P$ in Section~\ref{s:probdiff}, and also
state our Theorem. In Section~\ref{s:covariant} we describe how to
compute $\Delta P$ and $\Delta Q$ for the special case of
measurements, which are covariant with respect to phase space
translations, and show how to obtain the best constant $C$ in this
restricted class. Finally, in Section~\ref{s:compact} we show that
general measurements $M$ never outperform the covariant ones,
i.e., the bounds previously established also hold for joint
measurements without assuming any covariance condition.  Some
related ideas and versions of uncertainty will be discussed in the
last Section~\ref{s:other} of the paper.

%%%%%%%%%%%%%%%%%%%%%%%%%%%%%%%%%%%%%%%%%%%%%%%%%%%%%%%%%%%%%%%%%%%%%%
\section{Distance of observables on a metric space}
\label{s:probdiff}

\subsection{Monge distance of probability measures}
Let us fix some notation. If $X$ is some measurable space (i.e., a
space equipped with a $\sigma$-algebra of `measurable sets') a
probability measure $\mu$ on $X$ assigns to each measurable set a
probability in a countably additive way. Equivalently, we can
consider the expectation value functional induced by $\mu$, i.e.,
$\mu(f)=\int\!\mu(dx)\;f(x)$, where $f$ is any bounded measurable
function $f:X\to\Rl$. This functional, which we will denote by the
same letter $\mu$ is also called the Radon measure associated with
$\mu$. Whether a measure is primarily seen as a function on sets
or as a linear functional is largely a matter of taste. The Radon
measure point of view will have advantages in
Section~\ref{s:compact}, where we need to discuss measures with
non-zero weight at infinity. Hence we will use it throughout the
paper.  By $\delta_x$ we denote the point measure at $x\in X$,
i.e., $\delta_x(f)=f(x)$ for all $f$.

A natural way of describing the difference between two probability
measures $\mu_1$ and $\mu_2$ is to take the largest difference in
probabilities they can assign to any event, i.e., (up to a conventional
factor $2$):
\begin{eqnarray}\label{totvar}
  \norm{\mu_1-\mu_2}_1&=&2\sup_\sigma\abs{\mu_1(\sigma)-\mu_2(\sigma)}
  \nonumber\\
             &=&\sup_{\abs f\leq1}\abs{\mu_1(f)-\mu_2(f)}\;,
\end{eqnarray}
where the first supremum is over all measurable sets, and the
second is over all measurable functions with $\abs{f(x)}\leq1$ for
all $x$. This quantity is known as the norm difference with
respect to the norm of ``total variation''.

However, this distance between probability measures is totally useless
for defining a quantity like $\Delta Q$. As a measurable space $X=\Rl$
and $X=\Rl^2$ are isomorphic, so this structure knows nothing of the
topology of $X$, and of the closeness of points in $x$. For example take
two point measures $\delta_{x_1}$ and $\delta_{x_2}$ for distinct points
$x_1,x_2\in X$. Since there is a measurable set containing $x_1$ but not
$x_2$, we always have $\norm{\delta_{x_1}-\delta_{x_2}}_1=2$, even if the
points are ``very close'' and so the two point measures describe
practically indistinguishable probability distributions.

In order to set up a quantitative notion of the distance of
probability measures, according to which nearby point measures
would be close, too, we must have a notion of closeness for points
to begin with. Therefore, we fix a metric $d$ on $X$. The only
technical requirement linking the metric and the measurable
structure is that all continuous functions for the metric are
measurable. The idea is then to define the distance between
probability measures as the largest difference of expectation
values on ``slowly varying functions''.

\begin{defi}\label{def:dmu}
Let $X$ be a metric space with metric $d$. We define the
{\bf Lipshitz ball} $\Lambda$ of $(X,d)$ as the set of bounded
functions $f$ such that
\begin{equation}\label{lipsh}
  \abs{f(x)-f(y)}\leq d(x,y)\;,
\end{equation}
for all $x,y\in X$. Then, for any two probability measures $\mu_1,\mu_2$
on $X$  we define the distance as
\begin{equation}\label{dist}
  d(\mu_1,\mu_2)=\sup_{f\in\Lambda}\abs{\mu_1(f)-\mu_2(f)}\;.
\end{equation}
\end{defi}

\noindent Strictly speaking, it is another abuse of notation to
use the same letter for the metrics on points and on measures.
However, the two are very closely related. For example, if we take
two point measures, we find $d(\delta_{x},\delta_{y})=d(x,y)$,
where the inequality ``$\leq$'' follows by definition of the
Lipshitz ball, and the reverse inequality follows by observing
that the function $f(z)=d(x,z)$ (with a suitable cutoff to make it
bounded) is in $\Lambda$ by the triangle inequality.

There is an alternative ``dual'' definition of this distance,
going back to a problem by G. Monge\cite{Monge} in 1781. Consider,
instead of two probability distributions, two heaps of soil of
equal volume and the task of transforming one heap into the other
by moving around small amounts of soil. Suppose that for each such
move we have to pay a price proportional to the amount and to the
distance. Then the lowest possible price for the transformation is
called the {\it Monge distance} between $\mu_1$ and $\mu_2$. To
make this definition explicit, suppose we note for each bit of
soil the initial and final location. This will result in a
probability measure $\mu_{12}$ on $X\times X$, whose marginals are
the given measures $\mu_1$ and $\mu_2$, respectively. The price
payed will be proportional to
\begin{equation}\label{e:Mongeprice}
  D(\mu_{12})=\int\!\mu_{12}(dx,dy)\;d(x,y)\;.
\end{equation}
Clearly, for any $f\in\Lambda$, we will have
\begin{equation}\label{e:fmu12}
  \mu_1(f)-\mu_2(f)=\int\!\mu_{12}(dx\,dy)\ \bigl(f(x)-f(y)\bigr)
        \leq D(\mu_{12})\;.
\end{equation}
Then the supremum of the left hand side is the distance defined in
Definition~\ref{def:dmu}, whereas the infimum of the right hand
side is Monge distance. Due to a 1942 paper of L. Kantorovich
\cite{Kantor} the two are, in fact, the same. This result is very
much in the spirit of modern duality theory of convex optimization
problems. Duality also helps to understand the structure of
maximizing functions $f$ and minimizing joint distributions
$\mu_{12}$, which tend to be supported on the graph of a function,
provided the measures $\mu_1$ and $\mu_2$ are not too lumpy.
Uniqueness can be enhanced\cite{Caff} by replacing $d(x,y)$ in the
objective functional by $d(x,y)^{1+\epsilon}$, thereby putting an
extra penalty on scattering mass, and then letting $\epsilon\to0$.

Position and momentum take their values in a {\it vector space},
so we will briefly note some special properties of the Monge
metric in this case. We assume that the metric $d$ is consistent
with the linear structure, namely translationally invariant, and
homogeneous with respect to scaling. In other words, we require
$d(x,y)=\abs{x-y}$ for some vector space norm $\abs\cdot$. The
scaling property is important so that we can assign to distances
the same physical units as to the coordinates. Another key
operation that requires the vector space structure is adding noise
from an independent source. On the level of probability measures
this is represented by the convolution $\mu\ast\nu$:
\begin{equation}\label{e:convol}
  (\mu\ast\nu)(f)=\int\!\!\mu(dx)\int\!\!\nu(dy)\;f(x+y)\;.
\end{equation}
Basic properties of the metric on probability measures are summarized in
the following Lemma.

\begin{lemma}\label{lem:dvector} Consider $X=\Rl^n$, with a metric $d$ given by
a vector space norm.
\begin{enumerate}
\item Then for any probability measures $\mu,\nu$, we have the
      inequality\newline
      $d(\mu,\mu\ast\nu)\leq\int\!\nu(dy)\ \abs y$.
\item Let $n=1$, and let $F_i(t)=\mu_i\bigl((-\infty,t]\bigr)$
      denote the distribution functions of two probability measures on $\Rl$.
      Then\newline
      $d(\mu_1,\mu_2)=\int_{-\infty}^\infty\!\!dx\;\abs{F_1(x)-F_2(x)}$.
\end{enumerate}
\end{lemma}

\noindent The first estimate follows by inserting for $\mu_{12}$
in (\ref{e:fmu12}) the joint distribution  of $x$ and $x+y$
implied by the independence of $x$ and $y$ according to
(\ref{e:convol}). For the second statement note that we can take
the supremum over $f\in\Lambda$ over the subset of piecewise
differentiable functions with $\abs{f'(x)}\leq1$ such that $f'$
has compact support, and write
\begin{equation}\label{e:dpartialint}
 \mu_i(f)=-\int_{-\infty}^\infty\!\!\!\!dx\ F_i(x) f'(x) \;,
\end{equation}
up to boundary terms which cancel in the difference
$\mu_1(f)-\mu_2(f)$. This also provides a formula for the
maximizing $f$: we take $f'(x)=\pm1$, depending on the sign of
$F_1(x)-F_2(x)$.

%%%%%%%%%%%%%%%%%%%%%%%%%%%%%%%%%%%%%%%%%%%%%%%%%%%%%%%%%%%%%%%%%%%%%%
\subsection{Distance of Observables}

Let us now consider observables over $X$, i.e. quantum devices,
which produce an output $x\in X$ in every single experiment. Let
us take the quantum particles to be described in a Hilbert space
$\HH$, so that every preparation of quantum particles is described
by a density operator $\rho$. For any such preparation $\rho$, the
outputs of the device are then distributed with respect to a
probability measure $\mu_\rho$ on $X$. Since the map $\rho\mapsto
\mu_\rho(f)$ is affine in $\rho$ and bounded, for every bounded
measurable function $f$, there is an operator
$\Fobs(f)\in\BB(\HH)$ such that
\begin{equation}\label{e:F}
   \mu_\rho(f)=\tr(\rho\; \Fobs(f))\;,
\end{equation}
Then $\Fobs$ is a linear operator, taking positive functions to
positive operators, and $\Fobs(1)=\idty$. Evaluated just on the
indicator functions we get a positive operator values measure
(POVM), from which the values for general $f$ are recovered by
integration: We have $\Fobs(f)=\int\!\Fobs(dx)\;f(x)$. Either the
measure or the linear operator $\Fobs$ will called an observable,
and the two are denoted by the same letter. Of course, an
important special case is that each value of the measure is a
projection, which is equivalent to $\Fobs(fg)=\Fobs(f)\Fobs(g)$.
Such observables will be called {\it projection valued} (PVM).

How should we define the distance of observables now? An approach
based on joint distributions is not feasible, because very often
positive operator valued measures do not admit an extension to a
joint observable, so the measured outputs of two observables
typically cannot be seen in the same experiment and compared for
each single shot separately. We can, however, compare
distributions. And since equality of observables $\Fobs_1,\Fobs_2$
means, by definition, that the probability measures $\mu_{1,\rho}$
and $\mu_{2,\rho}$ coincide for every $\rho$, it is natural to say
that two observables are similar if they give similar probability
distributions for all states, in the sense of the metric defined
previously. Hence we set, for any two observables
$\Fobs_1,\Fobs_2$ on the same metric output space $(X,d)$ and for
quantum systems with the same Hilbert space $\HH$:
\begin{eqnarray}\label{e:dFF}
  d(\Fobs_1,\Fobs_2)
     &=&\sup_\rho\sup_{f\in\Lambda}\abs{\tr\;(\rho(\Fobs_1(f)-\Fobs_2(f))}
  \nonumber\\
     &=&\sup_{f\in\Lambda}\norm{\Fobs_1(f)-\Fobs_2(f)}\;,
\end{eqnarray}
where the supremum over $\rho$ in the first line is over all
density operators, and in the second line we used that, for
hermitian operators $A$ we can express the operator norm as $\norm
A=\sup_\rho\abs{tr(\rho\;A)}$. Thus
''$d(\Fobs_1,\Fobs_2)\leq\varepsilon$'' is synonymous with the
bound $\abs{\tr(\rho\,\Fobs_1(f))-\tr(\rho\,
\Fobs_2(f))}\leq\varepsilon$, on differences of expectation
values, valid for all states $\rho$, and all bounded functions $f$
with Lipshitz slope at most $1$.

It is important to note that the dual characterization of the
metric as Monge distance cannot be transferred from the case of
scalar probability measures to the operator valued case. Of
course, for any fixed $\rho$ we get a joint probability
distribution $\mu_{12,\rho}$ minimizing cost for the Monge problem
of $\mu_{1,\rho}$ and $\mu_{1,\rho}$. However, in contrast to the
$\mu_{i,\rho}$, the function $\rho\mapsto\mu_{12,\rho}$ is not
affine in $\rho$, and hence there is no observable $\Fobs_{12}$
such that $\mu_{12,\rho}(f)=\tr(\rho\Fobs_{12}(f))$, and even if
there happens to be some joint measurement $\Fobs_{12}$ with
marginals $\Fobs_i$, providing an affine family of joint
distributions, this gives only a loose upper bound
$d(\Fobs_1,\Fobs_2)\leq\norm{\Fobs_{12}(d)}$.

%%%%%%%%%%%%%%%%%%%%%%%%%%%%%%%%%%%%%%%%%%%%%%%%%%%%%%%%%%%%%%%%%%%%%%
\subsection{Statement of the Theorem}

Let us now consider a  quantum mechanical system with $\degf$
canonical degrees of freedom, described in a Hilbert space $\HH$.
That is there are self adjoint operators $P_\mu,Q_\mu$,
$\mu=1,\ldots,\degf$ satisfying the canonical commutation
relations
\begin{eqnarray}\label{e:ccr}
  i[P_\mu,Q_\nu]&=&\hbar\delta_{\mu\nu}\idty  \nonumber\\
  i[P_\mu,P_\nu]&=&i[Q_\mu,Q_\nu]=0
\end{eqnarray}
on a dense set of vectors, on which all real linear combinations
of these operators are essentially self-adjoint. Note that we do
admit additional degrees of freedom unrelated to the $P_\mu,Q_\mu$
under consideration. Under these conditions there are joint
spectral measures for the $Q_\mu$, i.e., there is a unique
projection valued observable on $X=\Rl^f$ such that
$Q_\mu=\int\!\Qobs(dx)x_\mu$. As a metric on $X$ we take some
metric derived from a norm $\abs\cdot$ on position space, such as
the Euclidean metric when we consider a single particle. By $\abs
Q$ we denote the operator
\begin{equation}\label{e:absQ}
  \abs Q=\Qobs(\abs\cdot)=\int\!\Qobs(dx)\;\abs x\;.
\end{equation}
Similarly, we consider the  momentum observable $\Pobs$, which is the
joint spectral measure of the $P_\mu$, and choose a suitable norm on
momentum space.

%\begin{theorem}
\vskip16pt\noindent{\bf Theorem.}\quad{\em
 Let $\Qobs,\Pobs$ be the position and momentum
observables of a system with $\degf$ degrees of freedom in a
Hilbert space $\HH$.
 Let $\Mobs$ be an observable on $\Rl^\degf\times\Rl^\degf$ with marginals
 $\Mobs_1$ and $\Mobs_2$.
Then
\begin{equation}\label{e:MUR}
  d(\Qobs,\Mobs_1)\cdot d(\Pobs,\Mobs_2)\geq C\hbar\;.
\end{equation}
The best constant $C$ in this inequality is determined as
 $C\hbar={E_0^2}/{4ab}$, where $E_0$ is the lowest eigenvalue
of the operator
\begin{equation}\label{e:Hlow}
  K=a\abs Q+b\abs P
\end{equation}
for some positive constants $a,b>0$. Equality in (\ref{e:MUR})
holds for a suitable covariant observable.
 }\vskip5ptplus30pt
%\end{theorem}

Of course, the numerical value of the constant $C$ does depend on
the two metrics chosen, and on the number $\degf$ of degrees of
freedom. For a single degree of freedom, with $\abs\cdot$ the
usual absolute value we get
\begin{equation}\label{e:Cfor1}
    C\approx0.304745
\end{equation}
The unique covariant observable attaining this bound is determined
(numerically) in Section~\ref{sec:opticov}. It is not equal to the
covariant observable based on coherent states, which realizes the
uncertainty product $C'=1/\pi\approx0.3183$.

%%%%%%%%%%%%%%%%%%%%%%%%%%%%%%%%%%%%%%%%%%%%%%%%%%%%%%%%%%%%%%%%%%%%%%
\subsection{Weight at infinity}\label{s:compactify}

Before going into the proof of the Theorem, we have to be a bit
more precise about the class of functions $f$, for which we need
expectations $\mu(f)$ and operators $\Fobs(f)$ to be defined. This
subsection is somewhat technical, and can be skipped by those who
are only interested in the construction of joint measurements
saturating the bound.

The issues we discuss in this subsection do not require
covariance, and make sense for a general locally compact metric
space $(X,d)$, in our case either phase space, position space or
momentum space. Integration on locally compact spaces can be
developed nicely as a theory of linear functionals (``Radon
measures'') on the space $\Ccs(X)$ of continuous functions of
compact support. This approach is advocated e.g., by
Bourbaki\cite{Bou} (see also Dieudon\'e\cite{Dieu} for the simpler
special case of metrizable separable spaces, which is all we need
here).

However, we need expectation values not just for $f\in\Ccs(X)$.
For example, in order to define the normalization of probability
measures we need to integrate the function ``1''. A large part of
measure and integration theory is devoted to extending the
definition of integrals to larger and larger classes of functions.
For the normalization one defines
\begin{equation}\label{e:mu1}
  \mu(1)=\sup\bigl\{\mu(f)\bigm|\; f\in\Ccs(X),\quad  f\leq1\bigr\}\;.
\end{equation}
The extension of $\mu(f)$ to all bounded measurable functions $f$
follows similar limit processes. For our purposes, however, it is
only necessary to evaluate expectations on $f\in\Lambda$, so we
can compute the distance of probability measures via
Definition~\ref{def:dmu}. Therefore we will stay with the minimal
space of functions necessary for that purpose, which is the
C*-algebra generated by $\Lambda$, the algebra $\Cuc(X)$ of
bounded uniformly continuous functions on $X$.

Now in Section~\ref{s:compact} we will construct directly some
normalized positive linear functionals (i.e., ``states'') $\mu$ on
$\Cuc(X)$, and we would like to conclude that such a $\mu$ defines
a measure on $X$. The problem is, however, that since
$1\in\Cuc(X)$, an equation like (\ref{e:mu1}) is now no longer a
definition of the left hand side, but a property of the functional
$\mu$. And we will see that it may indeed fail to be true. In
other words, for such functionals the monotone convergence theorem
$\sup_n\mu(f_n)=\mu\left(\sup_nf_n\right)$ for the pointwise
supremum of functions on $X$ may fail.

How does this fit in with the equivalence between `measures as set
functions' and `measures as linear functionals' proclaimed at the
beginning of this subsection and in Section~\ref{s:probdiff}? This
can be understood by considering the example of the algebra ${\cal
A}$ of bounded continuous functions on the open unit disk, which
have a continuous extension to the closed disk. As an algebra this
is identical with the continuous functions on the closed disk, but
what is meant by `pointwise supremum' now depends on what we
consider as the domain of these functions. For example, a point
measure on the boundary has the property of giving zero
expectation to any function with compact support inside the open
disc, producing a failure Eq.~(\ref{e:mu1}). It is clear from this
example that the equivalence between measures as set functions and
as linear functionals may require that we suitably extend the
underlying space, i.e., we consider measures on the closed disc
rather than just the open disc.

The general situation for a locally compact metric space $X$ is
quite similar to this example. Like every commutative C*-algebra,
the algebra $\Cuc(X)$ is isomorphic to the continuous functions
${\cal C}(\widehat X)$ on a compact space $\widehat X$, called the
{\it Gelfand spectrum} of $\Cuc(X)$. $\widehat X$ can be
constructed as the set of pure states on the algebra $\Cuc(X)$.
Since evaluation at a phase space point is a pure state, we have
$X\subset\widehat X$, and $\widehat X$ is a {\it compactification}
of $X$. The additional points $\widehat X\setminus X$ should be
thought of as points at infinity, and clearly a measure may be
supported on such points so that the restriction of $\mu$ to
$\Ccs(X)$ is zero, and Equation~(\ref{e:mu1}) is violated as
$1\neq0$. The points at infinity have a very rich
structure\cite{PHyUni}, but in this paper we are only interested
in their collective weight with respect to a probability measure,
which is simply the difference between left and right hand side of
Equation~(\ref{e:mu1}): For the overall {\it weight at infinity}
of a Radon probability measure $\mu$ on $\Cuc(X)$ we introduce the
notation
\begin{equation}\label{e:muinfinity}
  \mu(\infty)=1-
   \sup\bigl\{\mu(f)\bigm|\; f\in\Ccs(X),\quad  f\leq1\bigr\}\;.
\end{equation}

For a positive operator valued measure we can take exactly the
same definition: the supremum exists in the weak operator
topology, because the net of functions $f$ is directed.
Equivalently, we can apply the scalar definition to every measure
$\mu_\rho(f)=\tr(\rho \Mobs(f))$, and define the operator weight
at infinity by  $\tr(\rho \Mobs(\infty))=\mu_\rho(\infty)$ for
every $\rho$. The key observation, allowing us later to eliminate
weights at infinity, is the following

\begin{lemma}\label{l:sameinfty}\
\begin{enumerate}
\item For any Radon probability measures $\mu_1,\mu_2$ on a
locally compact metric space $(X,d)$: \quad
  $d(\mu_1,\mu_2)<\infty$ implies
    $\mu_1(\infty)=\mu_2(\infty)$.
\item Let $\Mobs$ be an observable on phase space, whose marginals have
finite distance to the standard position and momentum observables,
respectively.  Then $\Mobs(\infty)=0$.
\end{enumerate}
\end{lemma}

\begin{proof}
As a a net of functions $f_R\in\Ccs(X)$ we choose
\begin{equation}\label{e:fR}
  f_R(x)
    =\left\lbrace\begin{matrix}
        1-d(0,x)/R&\qquad\mbox{if}\ &d(0,x)\leq R\\
        0         &\qquad\mbox{if}\ & d(0,x)>R
     \end{matrix}\right.
\end{equation}
where $0\in X$ is an arbitrarily chosen reference point. Since a
locally compact metric space is the union of the compact balls
$\{x|\,d(0,x)\leq R\}$ this family eventually dominates every
function $f\in\Ccs(X)$ with $f\leq(1-\varepsilon)$. Hence
$1-\mu(\infty)=\lim_{R\to\infty}\mu(f_R)$ for every probability
measure. Then, since $Rf_R\in\Lambda$,
\begin{equation}
  \abs{\mu_1(f_R)-\mu_2(f_R)}\leq\textstyle\frac1R\;d(\mu_1,\mu_2)
\end{equation}
and the first result follows by taking the limit ${R\to\infty}$.

For the second statement consider the inequality
\begin{eqnarray}\label{e:cutoff2}
  1-f_R(p)f_R(q)
       &=&(1-f_R(p))+f_R(p)\bigl(1-f_R(q)\bigr)
           \nonumber\\
      &\leq& (1-f_R(p))+(1-f_R(q))\;,\nonumber
\end{eqnarray}
apply $\Mobs$ and take the limit $R\to\infty$ to get
$\Mobs(\infty)\leq\Mobs_1(\infty)+\Mobs_2(\infty)$, where
$\Mobs_i$ are the two marginals. But since the standard position
and momentum observables have zero weight at infinity, part 1 of
the Lemma shows that $\Mobs_i(\infty)=0$.
\end{proof}

%%%%%%%%%%%%%%%%%%%%%%%%%%%%%%%%%%%%%%%%%%%%%%%%%%%%%%%%%%%%%%%%%%%%%%
\section{Covariant observables}\label{s:covariant}
\subsection{Phase space covariant observables}

It is not a priori clear that there exist approximate joint
measurements of $P$ and $Q$, making $\Delta Q=d(\Qobs,\Mobs_1)$
and $\Delta P=d(\Pobs,\Mobs_2)$ finite. But there is a simple, and
even well-known construction for joint measurements of position
and momentum achieving just that. These phase space observables
have the additional property, that the unitary groups of
translation (generated by the momentum operators) and boosts
(generated by the position operators) act like a shift in phase
space on the arguments of $\Mobs$. Let us introduce the Weyl
operators (phase space translations)
\begin{equation}\label{e:weyl}
  W(p,q)=\exp\frac i\hbar\left(q\cdot P-p\cdot Q\right)\;.
\end{equation}
We will assume in this section that beyond the canonical ones
under consideration there are no additional degrees of freedom,
i.e., the Weyl operators act irreducibly on $\HH$. Then by von
Neumann's Uniqueness Theorem\cite{vNeu} we can take, up to unitary
equivalence, $\HH=\LL2(\Rl^\degf,dx)$, $Q_\mu$ the multiplication
by the coordinate $x_\mu$ and $P_\mu=\frac\hbar
i\frac\partial{\partial x_\mu}$. The Weyl operators in this
representation become
\begin{equation}\label{e:weylrep}
  (W(p,q)\psi)(x)=\exp\frac i\hbar\left(-\frac{p\cdot q}2-p\cdot x\right)
                  \ \psi(x+q)\;.
\end{equation}

Then, denoting by $(\tau_xf)(y)=f(y-x)$ the translate of a function on a
vector space by $x$, we get the shift covariance property of the standard
position observable becomes
\begin{equation}\label{e:Qcovariant}
  \Qobs(\tau_qf)=W(p,q)^*\Qobs(f)W(p,q)\;,
\end{equation}
 for all bounded measurable $f$ and all $p,q\in\Rl^\degf$. There is an
analogous property of $\Pobs$, and we define a {\it covariant phase space
observable} by the equation
\begin{equation}\label{e:covariance}
  \Mobs(\tau_{(p,q)}f)=W(p,q)^*\Mobs(f)W(p,q)\;.
\end{equation}
It turns out that there is a closed formula for all such
observables, described in the following Lemma. Recall from the
previous section that $\Cuc(X)$ denotes the algebra of bounded
uniformly continuous functions on phase space $X$, and $\Ccs(X)$
the subalgebra of functions with compact support. The Lemma is
well known, and versions of it can be found in
,PHyUni\cite{Holevo-old,Davies,QHA}.

\begin{lemma}\label{lem:covobs} Let $M$ be a covariant observable on phase space,
i.e., a linear map $\Mobs:\Cuc(X)\to\BB(\HH)$, taking positive
functions to positive operators, and satisfying
Eq.~(\ref{e:covariance}). Suppose that $\Mobs$ has zero weight at
infinity, i.e., $\sup\{M(f)|f\in\Ccs(X),\;f\leq1\}=\idty$. Then
there is a positive operator $\Mdens$ with $\tr(\Mdens)=1$ such
that
\begin{equation}\label{e:covobs}
  \Mobs(f)=\int\!\!\frac {dp\;dq}{(2\pi\hbar)^\degf}\;f(p,q)\;
            W(p,q)^*\Mdens\, W(p,q)\;.
\end{equation}
Conversely, this formula defines a covariant observable for every
$\Mdens$. The integral is a weak integral, i.e., for every density
operator $\rho$ we have to compute the expectation as
\begin{equation}\label{e:covobsweak}
  \tr\bigl(\rho\;\Mobs(f)\bigr)
        =\int\!\!\frac {dp\;dq}{(2\pi\hbar)^\degf}\;f(p,q)\;
            \tr\bigl(\rho\;W(p,q)^*\Mdens\, W(p,q)\bigr)\;.
\end{equation}
\end{lemma}

Here the operator $\Mdens$ is appropriately called a density
operator for two separate reasons: on the one hand as positive
operator of trace 1, and on the other hand as the ``Radon-Nikodym
derivative'' of the observable at the origin. The fact that the
trace under the integral is an integrable function, and, in fact a
probability density on phase space follows from the fundamental
``square integrability'' property of Weyl operators. In the
special case that $\Mdens$ is a coherent state (ground state of an
oscillator Hamiltonian) this probability density is also known as
the Husimi function\cite{Husimi} of $\rho$.

It is now easy to compute the marginals of $\Mobs$. Of course, the
marginals $\Mobs$ will inherit a covariance property: For example,
the position-like marginal $\Mobs_1$ will have the same property
(\ref{e:Qcovariant}) as the position observable. Since each
$\Mobs_1(f)$ commutes with momentum translations, these operators
must be functions of position, and the covariance for position
shifts forces $\Mobs_1$ to be equal to $\Qobs$ up to some smearing
by convolution with a fixed probability density. Explicitly, we
get the required density from the form of the Weyl operators. The
result is
\begin{equation}\label{e:mQ}
    m^Q(f)=\tr(\parity^*\Mdens\parity\;\Qobs(f))\;,
\end{equation}
where $\parity$ is the parity operator
$(\parity\psi)(x)=\psi(-x)$. With the analogous expression for
momentum and the definition (\ref{e:convol}) of convolution we
then have
\begin{equation}\label{e:covmarginal}
  \Mobs_1=\Qobs\ast m^Q\quad\mbox{and}\quad
  \Mobs_2=\Pobs\ast m^P\;.
\end{equation}

To summarize: the marginals of a covariant phase space observable
can be simulated in the following way: one simply makes the
corresponding ideal position or momentum measurement, and adds
some noise from a source which independent of the quantum state.
The noise distributions $m^Q$ and $m^P$ are the position and
momentum distributions of a density operator (namely
$\parity^*\Mdens\parity$), hence there is the usual tradeoff: if
we insist on a good position measurement, i.e., sharply peaked
$m^Q$, then $m^P$ will be very spread out, and much noise is added
to momentum, and conversely. In the following section we make this
quantitative in the sense of the distance of observables.

\subsection{Optimizing over covariant observables}\label{sec:opticov}

\begin{figure}[ht]\label{hyperbolas}\centering
 \epsfxsize=5cm \epsffile{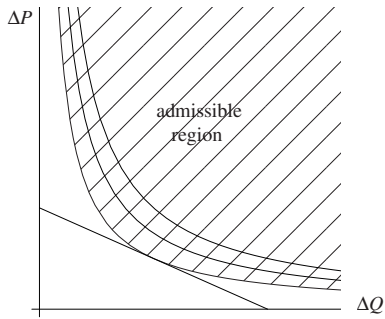}
 \caption{{\it Admissible region for pairs
     $(\Delta Q,\Delta P)=(\tr(\rho\abs\Qobs),\tr(\rho\abs\Pobs))$.
     The tangent shown is the contour line of $a\Delta Q+b\Delta P$
     realizing the minimum expectation of $K$.}}
 \end{figure}

The uncertainties $d(\Mobs_1,\Qobs)$ follow from
Eq.~(\ref{e:covmarginal}) and Lemma~\ref{lem:dvector}: we get
\begin{equation}\label{e:uncovQP}
  d(\Mobs_1,\Qobs)=\int\!\!m^Q(dx)\;\abs x=\tr(\Mdens \abs\Qobs)\;,
\end{equation}
and the analogous relation for $P$. Here the inequality ``$\leq$''
follows from Lemma~\ref{lem:dvector}.1, and equality follows from
the formula $d(\delta_y,\nu)=\int\nu(dx)\,d(x,y)$ and the
observation that there are states $\rho$ whose position
distribution is arbitrarily close to a point measure. We now have
to determine which combinations of two positive numbers
\begin{equation}\label{e:uncertpair}
  \Bigl(d(\Mobs_1,\Qobs),d(\Mobs_2,\Pobs)\Bigr)
    =\Bigl(\tr(\Mdens\,\abs\Qobs),\tr(\Mdens\,\abs\Pobs)\Bigr)  \nonumber
\end{equation}
can be obtained by varying the density operator $\Mdens$. Because
the coordinates depend linearly on $\Mdens$, this is a convex set
in the plane. Moreover, if one pair $(\delta_1,\delta_2)$ is
possible, then so is $(\delta_1',\delta_2')$ with
$\delta_i'\geq\delta_i$, because we can replace  $\Mdens$ by a
suitable average over translates, and can vary the distribution of
translation vectors from sharply concentrated to very broad. An
important point is that we can also apply the dilation symmetry
$Q_\nu\mapsto\lambda Q_\nu$, $P_\nu\mapsto\lambda^{-1} P_\nu$.
This is shown in Figure~\ref{hyperbolas}: with every point the
admissible region contains the entire hyperbola through that
point.
In order to find the parameter $C\hbar$ for the boundary
hyperbola, consider the lowest admissible expectation $E_0$ for a
linear combination
\begin{equation}\label{e:Hsum}
    K=a\abs\Qobs+b\abs\Pobs\;,
\end{equation}
with $a,b>0$. This is the same as the smallest $a\Delta Q+b\Delta
P$ with $(\Delta Q,\Delta P)$ in the admissible region of
Figure~\ref{hyperbolas}. Clearly this will be attained on the
boundary hyperbola $(\Delta Q)(\Delta P)=C\hbar$, which gives
$E_0=2\sqrt{abC\hbar}$. Solving for $C$ we find the statement of
the Theorem, for the special case of covariant $\Mobs$.

We still have to clarify whether the bound is attained, i.e.,
whether there is really an eigenvalue $E_0$ at the bottom of the
spectrum of $K$. This is equivalent to the same question for the
top of the spectrum of the operator $(K+\idty)^{-1}$, and we will
show it by verifying that $(K+\idty)^{-1}$ is a compact operator.
This also shows that the relevant eigenvalue has finite
multiplicity. A quick way to show compactness is by the
correspondence theory of Ref\cite{QHA,PHyUni}: we only need to
show that $(p,q)\mapsto W(p,q)^*(K+\idty)^{-1}W(p,q)$ is
continuous in norm (which is obvious by a resolvent equation), and
that the function
\begin{equation}\label{e:revolvent}
  k(p,q)=\bigl\langle W(p,q)\Phi,
   (K+\idty)^{-1}W(p,q)\phi\bigr\rangle
\end{equation}
goes to zero at infinity, when $\Phi$ is some fixed Gaussian wave
function. But since the operator inverse is decreasing with
respect to operator ordering, we have %\linebreak
$(K+\idty)^{-1}\leq(a|P|+\idty)^{-1}$, from which we get the
estimate $k(p,q)\leq {\rm const} (a|p\,|+1)^{-1}$, and similar
estimate for $q$. Hence $k$ goes to zero.

\begin{figure}[h]\label{bump}\centering
 \epsfxsize=7cm \epsffile{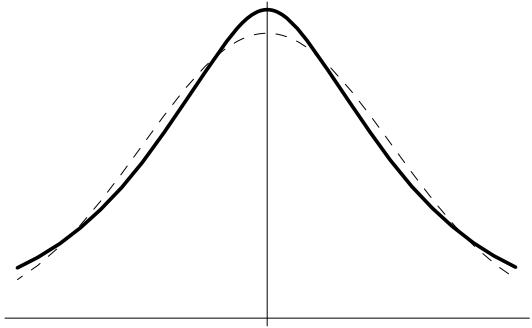}
 \caption{{\it The ground state wave function of the operator $K$
 (solid line), compared to the Gaussian oscillator ground state
 (dashed) for one degree of freedom. }}
 \end{figure}

In practice the computation of $E_0$ is best done by using the
symmetry of the problem. For several degrees of freedom and
Euclidean norms this is rotation symmetry, for one degree of
freedom, we still have reflection symmetry. In addition, it is
useful to take $a=b=\hbar=1$, and use the Fourier transform
symmetry of $K$. Since the $E_0$-eigenspace is finite dimensional,
we can seek joint eigenvectors of $K$ and the symmetries.  Then
$K$ is truncated to a subspace spanned by finitely many
eigenfunctions of the harmonic oscillator with fixed symmetry, and
the resulting matrix (which can be constructed symbolically, i.e.,
with infinite precision) is numerically diagonalized. One readily
finds that the ground state of $K$ is close to the oscillator
ground state (see Fig.~\ref{bump}), and is realized for angular
momentum $\ell=0$ (resp. even parity), and vectors invariant under
Fourier transform.
For more than one degree of freedom we get the following table

\vskip12pt

\begin{tabular}{c|c|c}
 dimension&$C$&$C'$\\ \hline
 1& 0.3047 &$1/\pi$= 0.3183\\
 2& 0.7628 &$\pi/4$= 0.7853\\
 3& 1.2457 &$4/\pi$= 1.2732\\
 42& 20.710 & \hskip30pt20.751
 \end{tabular}

%%%%%%%%%%%%%%%%%%%%%%%%%%%%%%%%%%%%%%%%%%%%%%%%%%%%%%%%%%%%%%%%%%%%%%
\section{Reduction to the covariant case}\label{s:compact}

In this section we will prove that in order to construct a joint
measurement $\Mobs$ with small uncertainties we can restrict attention to
the covariant observables studied in the previous section. The basic idea
is to average over phase space translations, thus turning a given
observable $\Mobs$ into a covariant one $\Mobsc$ with at least as small
error bounds.

The basis for the construction is a so-called {\it invariant mean}
\cite{invmean} on the group of phase space translations: this associates
to any bounded continuous function $f$ on phase space a number $\eta(u)$
such that $u\mapsto\eta(u)$ is linear, positive on positive functions,
normalized as $\eta(1)=1$, and invariant in the sense that
$\eta(\tau_xu)=\eta(u)$. The existence of invariant means is by no means
obvious. Any constructive procedure based on integrating over larger and
larger sets, and dividing by the volume of the set, will be convergent
only for `well behaved' functions, such as almost periodic ones or
functions going to zero at infinity. The latter always average to zero,
i.e., as a set an invariant mean has weight at infinity $\eta(\infty)=1$
in the sense of Eq.~(\ref{e:muinfinity}). A functional $\eta$ defined on
{\it all} bounded continuous functions can indeed not be constructed
explicitly, and we know its existence only via the axiom of choice.

Now let $f$ be a uniformly continuous function on phase space, and $\rho$
a density operator. Then we define the operator $\Mobsc(f)$ by
\begin{eqnarray}\label{e:Mav}
    \tr(\rho\Mobsc(f))
      &=& \eta\left(u(\rho,f)\right)\;, \\ \mbox{where}\quad
  u(\rho,f)(p,q)&=&
           \tr\left(W(p,q)^*\rho\;W(p,q) \Mobs(\tau_{(p,q)}f)\right)
\nonumber
\end{eqnarray}
The function $u(\rho,f)$ is designed so that it is constant, if $\Mobs$
is covariant, in which case case $\Mobsc=\Mobs$. For arbitrary $\Mobs$ it
is still always uniformly continuous, because for uniformly continuous
functions $f$ translation is norm continuous, and because Weyl operators
are strongly continuous in $(p,q)$, making $(p,q)\mapsto
W(p,q)\rho\;W(p,q)^*$ continuous in trace norm. The function $u$ is also
bounded by the upper bound $\norm f_\infty$ for $f$. It follows that the
invariant mean $\eta$ is applicable, and that the right hand side
(\ref{e:Mav}) is a bounded linear functional on the convex set of density
operators, and, consequently, there is a unique bounded operator
$\Mobsc(f)$. Obviously, $\Mobsc(f)$ is linear in $f$, positive on
positive functions, and normalized ($\Mobsc(1)=\idty$). By invariance of
the mean it is also evident that it is a covariant observable.

The crucial point we have to establish now is that averaging does
not increase uncertainty. To this end, let us consider the set
$\MM(\delta_1,\delta_2)$ of  observables $\Mobs$ with
$d(\Qobs,\Mobs_1)\leq\delta_1$ and $d(\Pobs,\Mobs_2)\leq\delta_2$.
In other words, we take $\MM(\delta_1,\delta_2)$ as the set of
observables $\Mobs$ on phase space, such that for all density
operators $\rho_1,\rho_2$ and all Lipshitz functions
$f_1,g_1:\Rl^\degf\to\Rl$ with $f,g\in\Lambda$:
\begin{eqnarray}
   \tr(\rho_1\, \Mobs_1(f_1))-\tr(\rho_1\,\Qobs(f_1)) &\leq& \delta_1 \nonumber\\
   \tr(\rho_2\, \Mobs_2(f_2))-\tr(\rho_2\,\Pobs(f_2)) &\leq& \delta_2 \;.
\end{eqnarray}
Suppose an observable $\Mobs$ satisfies these bounds. Then these
relations remain true, if we replace $\rho$ by
$W(p,q)\rho\;W(p,q)^*$, and the functions $f,g$ by appropriately
shifted ones. The terms involving $\Pobs$ and $\Qobs$ are
unchanged by this, but the terms with $\Mobs$  become continuous
functions of $(p,q)$ of the kind $u(\rho,f)$ in (\ref{e:Mav}), to
which we may apply the invariant mean $\eta$. As a result we find
that
\begin{equation}\label{e:Mav-epsilon}
  \Mobs\in\MM(\delta_1,\delta_2) \Rightarrow
  \Mobsc\in\MM(\delta_1,\delta_2) \;.
\end{equation}
Hence without increasing the uncertainty bounds we may replace
$\Mobs$ by the covariant observable $\Mobsc$.

This reduces the problem of characterizing the admissible pairs
$(\delta_1,\delta_2)$ to Section~\ref{s:covariant}, except for two
issues. The first is that throughout that section we had assumed
the Weyl operators to act irreducibly, i.e., there were no further
degrees of freedom present. However, von Neumann's uniqueness
Theorem\cite{vNeu} asserts that any system of Weyl operators can
be decomposed into a direct sum of irreducible systems. Let
$p_\alpha$ be the projections onto the irreducible direct
summands, which by definition commute with all Weyl operators.
Then we claim that $\Mobs\in\MM(\delta_1,\delta_2)$ implies that
$\sum_\alpha p_\alpha\Mobs p_\alpha\in\MM(\delta_1,\delta_2)$. The
argument for this is averaging as before, but over the group of
unitaries of the form $U=\sum_\alpha u_\alpha p_\alpha$ with
$\abs{u_\alpha}=1$. Clearly, such a direct sum of observables lies
in $\MM(\delta_1,\delta_2)$ iff every summand does. On the one
hand, this means that additional degrees of freedom cannot
increase the set of admissible $(\delta_1,\delta_2)$, and on the
other hand it means that if such a pair is admissible for
irreducible Weyl systems, we can construct observables with this
bound for arbitrary systems as direct sums.

The second issue we have to address is that $\Mobsc$ comes out as
a Radon observable, i.e., we get an operator $\Mobsc(f)$ for every
bounded uniformly continuous function $f$, but such an observable
might have non-zero weight at infinity. In fact, it is typical for
constructions based on a compactness argument (such as our appeal
to the existence of invariant means) that one has to verify in the
end that the construction does not lead to a wild element of a
compactified space. For example, if we had omitted the Weyl
operators from the definition of $u$, we could have still obtained
some observable $\Mobsc$ from Eq.~(\ref{e:Mav}) by averaging. But
rather than getting a covariant observable we would have found the
observable $\widetilde\Mobsc(f)=\eta(f)\idty$ which has {\it only}
weight at infinity. This is the reason we had to discuss weights
at infinity in Section~\ref{s:compactify}. In fact all the hard
work was already done there: When $\delta_1$ and $\delta_2$ are
finite, $\Mobsc\in\MM(\delta_1,\delta_2)$, implies that
$\Mobsc(\infty)=0$ by  Lemma~\ref{l:sameinfty}.2, and therefore,
by Lemma~\ref{lem:covobs}, that we can construct $\Mobsc$ by
integration over a density operator, and compute and optimize the
uncertainties as in Section~\ref{sec:opticov}. This concludes the
proof of the Theorem.

%%%%%%%%%%%%%%%%%%%%%%%%%%%%%%%%%%%%%%%%%%%%%%%%%%%%%%%%%%%%%%%%%%%%%%
\section{Other Uncertainty Relations}\label{s:other}

Of course, all uncertainty relations are related. Some variants
that are closely related to the present paper will be briefly
commented in this section.

\subsection{Measurement as Preparation}

One way of reducing measurement uncertainty to preparation
uncertainty is the projection postulate: According to this
postulate the state of a system after measurement is an eigenstate
of the measured observable for that eigenvalue, which happened to
be the outcome of the measurement. Let us assume some approximate
version of this postulate holds for the approximate measurement
$Q'$ for the microscope (Of course, this restricts the
applicability of our argument). Then by conditioning on the
particular value of $q'$ obtained from the $Q'$-measurement, we
could understand the position measurement as a preparation of
states with position and momentum spreads $\approx\Delta Q$ and
$\Delta P$. The relation between $\Delta Q$ and $\Delta P$ would
then be just another special case of the Preparation Uncertainty
Relation. The $\Delta P$ here would be the total variance of the
momentum distribution after the measurement, i.e., not really that
part of momentum uncertainty introduced by the measurement itself.
It could be much smaller that the initial momentum spread. So this
reduction of measurement uncertainty to preparation uncertainty is
straightforward only if we know that the initial state has sharp
momentum.

\subsection{Variance of covariant observables}
The curious constant $.3047$ in the relation we prove is perhaps
not so strange if one notes that the same constant appears in the
preparation uncertainty, if we choose to quantify the spread
$\Delta Q$ of position not by the square root of a second moment,
but by an absolute first moment. In fact, this is the way the
constant was derived in Section~\ref{sec:opticov}. So it is
suggestive to look for an interpretation of $\Delta Q$ and $\Delta
P$ for measurement uncertainty, which would also bring the
constant to $\hbar/2$. (From talks I gave about the subject I know
some colleagues find fault with any other constant). This can be
done at the expense of the Kantorovich interpretation of the Monge
distance as a worst case difference of expectation values, by
using a cost function for transport, which grows quadratically
with the distance (also known as the Wasserstein 2-metric).

This is especially suggestive for the purely covariant case, in
which the marginals of joint measurements are equivalent to adding
noise from an external source: one can then simply take the second
moment of the noise. This approach has been suggested also by
Holevo.

\subsection{Ozawa's Approach}
In a series of recent papers,Ozawa\cite{Ozawa} M. Ozawa has also
studied the measurement aspect of uncertainty. For this he
considers measurements described as detailed couplings to an
environment. Then one can explicitly point out a selfadjoint
operator of the combined object-apparatus system which describes
the momentum after the measurement. The `perturbation of momentum'
by the measurement is then represented by the difference of the
momentum operators before and after the measurement interaction,
and quantified by the expectation of the square of this operator.

This is definitely a departure from the operational approach to
quantum mechanics, since this difference of non-commuting
operators is not accessible in the given experiment. Of course,
{\it any} operator represents an observable. But to find a device
measuring just this operator is a highly non-trivial task. In
contrast, in our approach only the statistics of measurements on
the joint measuring device itself enters.

Nevertheless, there are interesting aspects in Ozawa's approach.
In particular, his analysis applies to every input state
separately, whereas our figures of merit involve a supremum over
all input states. Further relationships remain to be clarified.


\begin{thebibliography}{99}
\bibitem{Hei}W. Heisenberg, \"Uber den anschaulichen Inhalt der
 quantentheoretischen Kinematik und Mechanik, (The perceptible content of
 the quantum theoretical kinematics and mechanics),
 {\it Z.Phys~\bf 43} (1927) 172--198

\bibitem{Ken}E.H. Kennard,
   {\it Z.Phys.\ \bf44} (1927)326

\bibitem{Rob}H.P. Robertson,
   {\it Phys.Rev.\ \bf34} (1929)163

\bibitem{Husimi} K. Husimi,
   {\it Proc. Phys. Math. Soc. Japan \bf22} (1940)264.

\bibitem{HolevoUncert} A.S. Holevo,
   %Covariant measurements and uncertainty relations.
   {\it Rep. Math. Phys. \bf 16} (1979) 385-400

\bibitem{Monge}G. Monge, M\'emoire sur la th\'eorie des d\'eblais et de remblais.
{\it Histoire de l'cad\'emie Royale des Sciences de Paris, avec
les M\'emoires de Math\'ematique et de Physique pour la m\^eme
ann\'ee}, pages 666--704, 1781.

\bibitem{Kantor}L. Kantorovich, On the translocation of masses. {\it C.R. (Doklady)
Acad. Sci. URSS (N.S.) \bf37} (1942)199--201

\bibitem{Caff}L.A.~Caffarelli, M. Feldman, and R.J.~McCann,
 %Constructing optimal maps for Monge's transport problem as a limit of strictly convex costs.
 {\it J. Amer. Math. Soc. \bf 15} (2002) 1--26

\bibitem{Bou}N. Bourbaki, \'El\'ements de Math\'ematique,
  Int\'egration, (Chap. III), Hermann, Paris 1952

\bibitem{Dieu}J. Dieudonn\'e, Foundations of Modern Analysis,
(Chapter 13), Academic Press, 1970

\bibitem{PHyUni}R.F. Werner,
  %Physical uniformities on the state space of non-relativistic quantum mechanics.
  {\it Found Phys \bf13} (1983) 859--881


\bibitem{vNeu} J. von Neumann,
  % Die Eindeutigkeit der Schr\"odingerschen Operatoren,
   {\it Math. Ann. \bf104} (1931), 570

\bibitem{Holevo-old} A.S. Holevo, {\it Probabilistic and statistical
   aspects of quantum theory}, (North-Holland, Amsterdam 1982)

\bibitem{Davies}E.B. Davies, {\it Quantum theory of open systems},
(Academic Press, London 1976)

\bibitem{QHA}R.F. Werner,
 %Quantum harmonic analysis on phase space
 {\it J. Math. Phys. \bf25} (1984) 1404--1411

\bibitem{invmean}E. Hewitt, K.A. Ross: {\it Abstract harmonic
analysis}, Vol.I, Chapter IV.\S\,17 (Springer, Berlin 1963)


\bibitem{Ozawa}M. Ozawa, {\it Phys. Rev. A \bf67} (2003), 042105,
{\it Phys. Lett. A \bf318} (2003)  21--29, {\it Phys. Lett. A
\bf320} (2004), 367--374, and quant-ph/0307057




\end{thebibliography}
\end{document}